# The Analysis of the Relationship Between Movement of European Airport Gravity Centres and the Establishment of Air Alliances

Vahidin Jeleskovic and Anastasia Blokhina[1]

## Abstract

This paper delves into the intricate relationship between the formation of air alliances and the shifts in airport and economic gravity centers across European countries during the period spanning 1990 to 2019. Employing descriptive analysis and the weighted mean center methodology, it explores the interplay between air passenger numbers and economic indicators, revealing a close and interdependent correlation between these factors. The study sheds light on the dynamic landscape of economic gravity centers, which experienced discernible shifts over time. However, it observes even more pronounced transitions in airport gravity centers. Statistical t-tests underscore significant differences in standard deviations when comparing pre- and post-air alliance periods for airport gravity centers. These disparities serve as a testament to the profound impact of airline alliances on the distribution of air traffic. These findings underscore the pivotal role of air alliances in reshaping the aviation landscape, and they beckon further investigation into their influence on broader economic transformations. Notably, this study pioneers the use of geographical means and standard deviations for rigorous statistical testing of economic hypotheses, signifying a significant contribution to the field.

---

[1] Humboldt Universität zu Berlin: vahidin.jeleskovic@hu-berline.de and anastasia.blokhina@student.hu-berlin.de

# Introduction

Over an extended period of time, human population has exhibited a consistent trend of congregating in urban centres. This phenomenon is primarily driven by the allure of diverse employment opportunities, access to educational institutions, and a robust healthcare infrastructure. However, it is imperative to acknowledge that the transportation network plays an indispensable role in facilitating this population concentration by enabling seamless commuting and travel. Notably, major airports tend to be strategically located within bustling metropolises, which, in most cases, already boast well-established economic ecosystems. These airports further augment these urban hubs by serving as vital transfer and cargo hubs.

Nonetheless, airlines frequently encounter a slew of challenges, including regulatory constraints imposed by national authorities, exorbitant operational costs, and logistical complexities associated with long-distance flight transfers. In response to these challenges, airlines have increasingly formed alliances with counterparts from foreign nations. A significant milestone in this realm was the inception of the Star Alliance in May 1997 (Star Alliance, 2023). This pioneering alliance united five major airlines from Germany, Scandinavia, the USA, Canada, and Thailand into a single network, thereby elevating the overall quality of service. Subsequently, a multitude of other air alliances emerged. These later alliances demonstrated remarkable achievements, with three major ones – Star Alliance, SkyTeam, and Oneworld – jointly accounting for the carriage of approximately 59% of all passengers and commanding a 70% share of the inter-continental market in 2008 (Bilotkach and Hüschelrath, 2012).

The foremost advantage arising from the formation of air alliances lies in the ability of airline companies to operate in foreign countries, circumventing the restrictions imposed by bilateral agreements. Furthermore, alliances enable the creation of more extensive routes replete with convenient connections and reduced travel durations. Another significant benefit pertains to cost savings achieved through the sharing of operational expenses and the realization of economies of scale. This allows companies to diversify their destination networks without necessitating additional investments in their aircraft fleet or airport representation (Lordan et al., 2015). Additionally, some regions, particularly those distant from the home country's origin or destination, may not be economically viable for an airline to operate individually (Gudmundsson and Rhoades, 2001).

As previously mentioned, major urban centres wield a magnetic pull for diverse reasons, establishing themselves as gravitational hubs for population concentration. In the context of aviation, the concept of an airport gravity centre within a country represents an aggregated midpoint, computed from the coordinates of various airports. The reliability and relevance of this data necessitate that it be weighted by socio-demographic metrics such as passenger numbers. This approach yields a weighted airport gravity centre point whose shifts over the years can be comprehensively analyzed.

In this study, we will employ classical descriptive analysis alongside the weighted mean centre methodology to investigate the interplay between the formation of air alliances and the migration of airport gravity centres within European countries. Our research inquiry concentrates on discerning the impact of air alliances on the redistribution of airport gravity centres in Europe. Furthermore, given the predominant focus of this paper on descriptive analytics concerning the movement of gravity centres over time, our objective is to present a comprehensive depiction of this transformation, underpinned by historical socio-economic data and empirical facts

Our methodological approach will encompass an initial presentation of classical descriptive analytics pertaining to key economic indicators, followed by the computation of weighted mean centre coordinates spanning the extensive period from 1990 to 2019, culminating in a discussion of potential underlying factors. Finally, we will subject the data to rigorous statistical scrutiny to determine whether the establishment of air alliances has indeed influenced the positioning of European airport gravity centers.

## Literature Review

To elucidate the interrelation between two distinct concepts, namely airport gravity centres and air alliances, it is essential to commence by delving deeper into their respective meanings. At its core, the notion of a "gravity center" evokes its fundamental physical definition – the hypothetical point within an object where its weight converges. From both a geographical and economic vantage point, it can be construed as a pivotal locale on a region's cartographic canvas, representing the average epicenter of economic activities. The application of this methodological framework bestows a more comprehensive insight into economic shifts, as it encapsulates the complete spatial dispersion of economic activities without amalgamating them across regions (Quah 2011).

This method has been extensively employed in numerous research papers, spanning both regional and global contexts. For instance, Quah's scholarly inquiry scrutinizes the movement of the gravity centre of economic activities worldwide, utilizing GDP data, and discerns disparities across different years (Quah 2011). The author's findings reveal that in 1980, the gravity centre was situated within the Atlantic region, but by 2008, it had migrated eastward to Europe. Furthermore, prognostications suggest a further eastward trajectory, positioning the economic gravity centre between India and China by 2050. Meanwhile, Raźniak et al. (Raźniak, Dorocki, and Winiarczyk-Raźniak 2020) observe shifts in gravity centres across various economic sectors globally, with a noticeable trend gravitating towards Asia, particularly in healthcare and information technology.

The principal driving forces behind these gravity centre displacements can be attributed to the economic growth experienced by developing nations and the escalating share of their production in global trade over time. This pattern is consistent across all sectors and economic activities; for instance, the share of developed nations in global GDP production diminished by nearly 40% between 1990 and 2015, while the contributions of developing countries in Asia, Africa, and Latin America continued to ascend (Tóth and Szép 2019). The migration of economic gravity centres has far-reaching implications for the global economic landscape. As developing nations amass greater prominence in world production and trade, the polarizing forces diminish, fostering a greater imperative for international cooperation. Emerging economies such as India and China are poised to exert transformative influence on the global economic order in the foreseeable future (Dobson 2009)

Notably, the aviation industry, encompassing airports and air transportation, plays a pivotal role, intimately connected to both trade, through cargo transport, and human mobility. Airport gravity centers, in this context, epitomize the average nexus of air traffic activities within a given region. Regrettably, there is a dearth of comprehensive research on this subject, particularly concerning the European region and its intersection with other contributing factors. Understanding shifts in airport gravity centres is crucial for unraveling the dynamics of economic gravity concerning urbanization, trade, and employment patterns

As previously alluded to, the formation of air alliances has wielded significant influence on the aviation industry, resulting in route adjustments and altering the prominence of key airports. Furthermore, it has left a discernible footprint on the distribution and migration of airport gravity centers, a facet that will be explored in greater detail in subsequent sections of this paper. Air alliance members engage in a spectrum of collaborative practices, predominantly revolving around asset sharing and service cooperation. Examples of asset sharing encompass code sharing, revenue sharing, wet leasing, shared IT infrastructure, insurance pooling, and shared facilities (Gudmundsson and Rhoades 2001). Cooperation extends to practices such as block space, franchising, joint services, and joint marketing.

The establishment of air alliances exerts a transformative impact on both the aviation sector and the broader economy. Collaboration engenders joint pricing strategies, thereby reducing fares within interline markets, fostering technological advancement, and enhancing traffic density. Additionally, owing to spillover effects, alterations in one market segment ripple across others, driving down domestic fares. This, in turn, augments consumer welfare and the overall surplus in comparison to fragmented industry participants (Brueckner 2001). Furthermore, the market influence wielded by member companies of air alliances tends to burgeon (Kuzminykha and Zufan 2014; Park and Zhang 2000).

The shifting of airport gravity centres serves as a barometer for economic and aviation industry transformations. Notably, the inception of air alliances constitutes a seminal force in this paradigm shift, which is anticipated to manifest in the relocation of geographic airport centers. Subsequent sections of this paper will present our own descriptive analysis results, further elucidating this intricate interplay.

## Classical Descriptive Analysis

Airport centres of gravity represent more than mere geographical coordinates; they are the culmination of meticulous calculations that involve the weighted amalgamation of locations predicated on various socio-economic factors. This research endeavors to scrutinize both airport and economic centres of gravity, employing Gross Domestic Product (GDP) in current US dollars and the total volume of air passengers carried, encompassing both domestic and international travelers on air carriers registered within the respective country, as weighting parameters. The temporal scope of this analysis encompasses the years 1990 to 2019, utilizing annually aggregated data sourced from the World Bank database, specifically the World Development Indicators section.

The selected countries for this analysis constitute pivotal transport hubs within Europe, predominantly comprising members of the European Union. This selection process was grounded in a ranking of the world's busiest airports according to the Port Authority of New York and New Jersey (Port Authority of New York and New Jersey 2023), thereby ensuring a representative cross-section of primary air traffic locations across diverse European regions. The countries included in this study encompass Austria, Belgium, Finland, France, Germany,

Greece, Ireland, Italy, Luxembourg, the Netherlands, Portugal, Spain, Sweden, Switzerland, Turkey, and the United Kingdom.

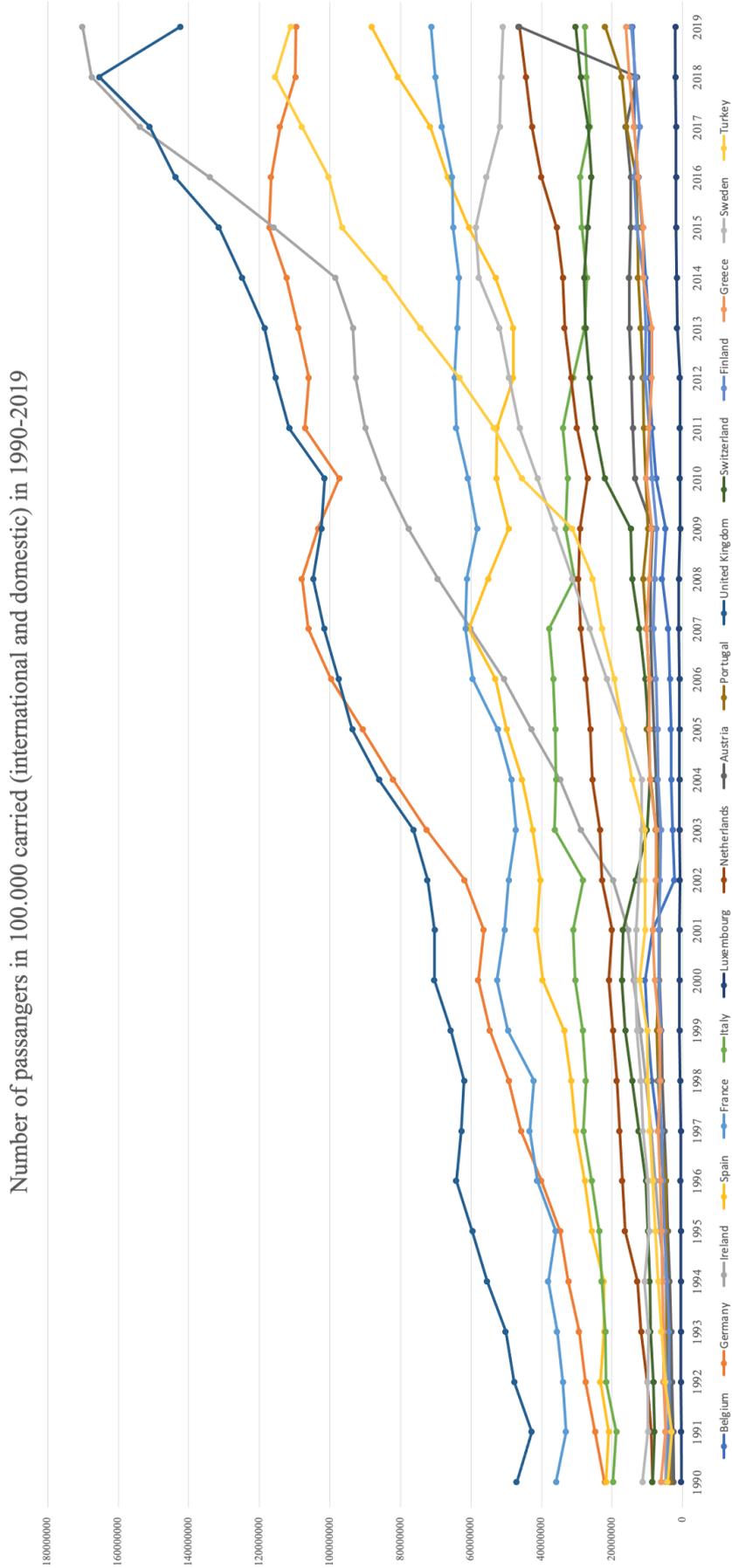

Figure 1. Number of passengers in 100.000 carried (international and domestic) in 1990-2019.

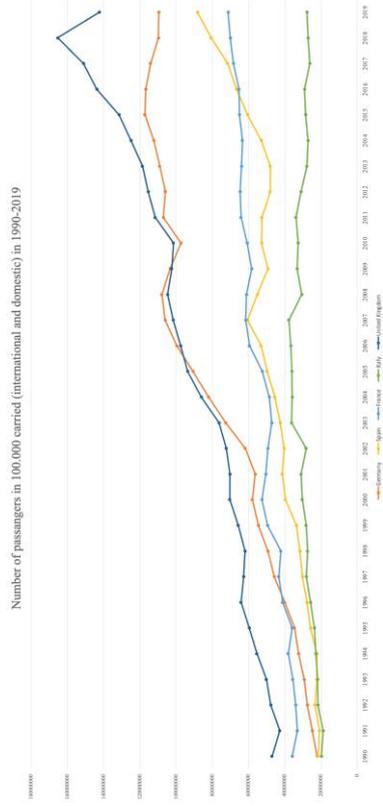
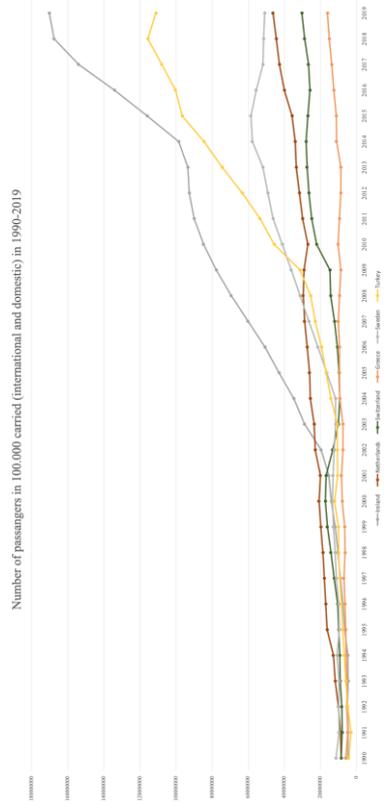
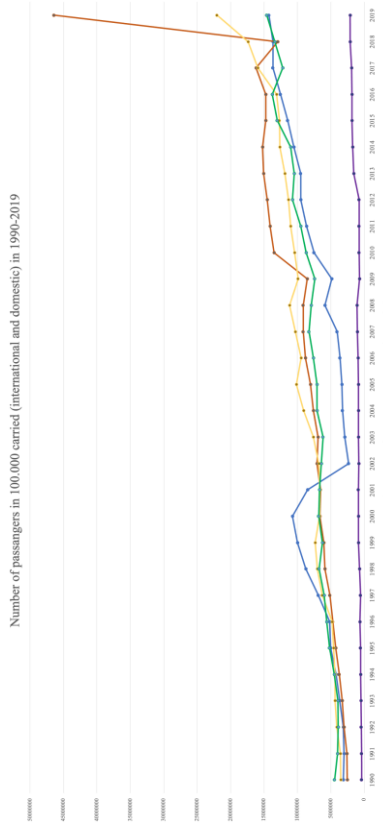
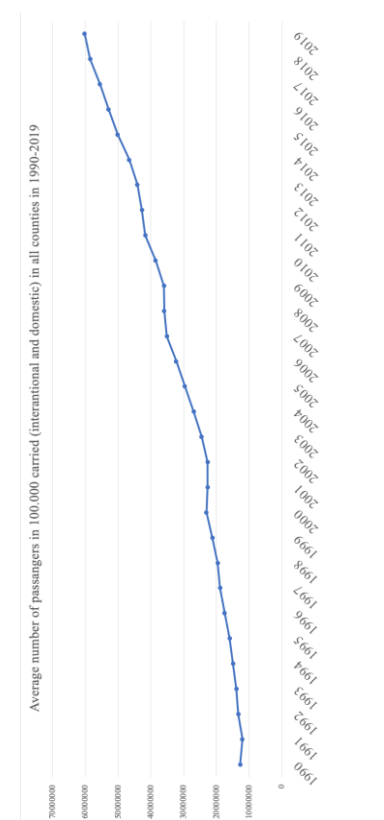

Figures 2-5. Number of passengers in 100.000 carried (international and domestic) in 1990-2019 and the average for all countries.

In order to enhance the representativeness of our descriptive analysis concerning the time-series data on passenger numbers, we have classified countries into three distinct groups based on a quantitative similarity factor. The initial group, characterized by the highest passenger figures, encompasses the United Kingdom, Germany, France, Spain, and Italy, as depicted in Figure 2. Evidently, these countries exhibit comparable trends in passenger number fluctuations over time. Notably, the United Kingdom and Spain demonstrate strikingly similar patterns, with a gradual increase in passenger numbers leading up to 2008, followed by a decline attributed to the Global Financial Crisis of 2008. Subsequently, both countries experienced rapid growth, with the United Kingdom peaking in 2018 and Spain in 2019.

Germany and France similarly demonstrate an upward trajectory in passenger numbers until 2019, with Germany exhibiting a steeper ascent. Conversely, Italy's graph exhibits a flatter trend, characterized by a marginal increase from 2003 to 2007, followed by a sustained decline.

Figure 3, on the other hand, provides an illustration of air passenger numbers in the Netherlands, Sweden, Switzerland, Turkey, Ireland, and Greece. Until 2001, these countries exhibited fairly uniform passenger figures, characterized by relative stability. However, commencing from 2002, Ireland and subsequently, in 2009, Turkey experienced notable surges in passenger numbers, reaching peaks in 2019 and 2018, respectively. The Netherlands, Sweden, and Switzerland also underwent a gradual increase in passenger numbers over the years. In contrast, Greece's passenger numbers exhibited modest fluctuations, remaining relatively constant until 2019, in comparison to the other nations. Surprisingly, not all countries in this group were affected negatively by the Global Financial Crisis of 2007. Switzerland, in particular, witnessed a decline, while other nations sustained growth or remained unaffected.

Turning our attention to Figure 4, this figure portrays air passenger numbers from 1990 to 2019 in Belgium, Luxembourg, Austria, Portugal, and Finland. Belgium initiated an upward trend in passenger numbers in 1996, followed by a local peak in 2000, only to experience a sudden decline in 2002. Subsequently, the country registered gradual growth. Finland and Portugal, in contrast, exhibited relatively stable and less fluctuating passenger numbers, demonstrating a consistent upward trajectory until 2019. Austria, on the other hand, experienced a modest increase in passenger numbers from 1995 to 2008, followed by a period of fluctuation, and then a dramatic ascent in 2018, ultimately culminating in a significant drop in 2020, attributed to the pandemic-induced lockdown. Luxembourg, similarly, demonstrated a trend characterized by overall constancy, with a slight increase observed until 2020.

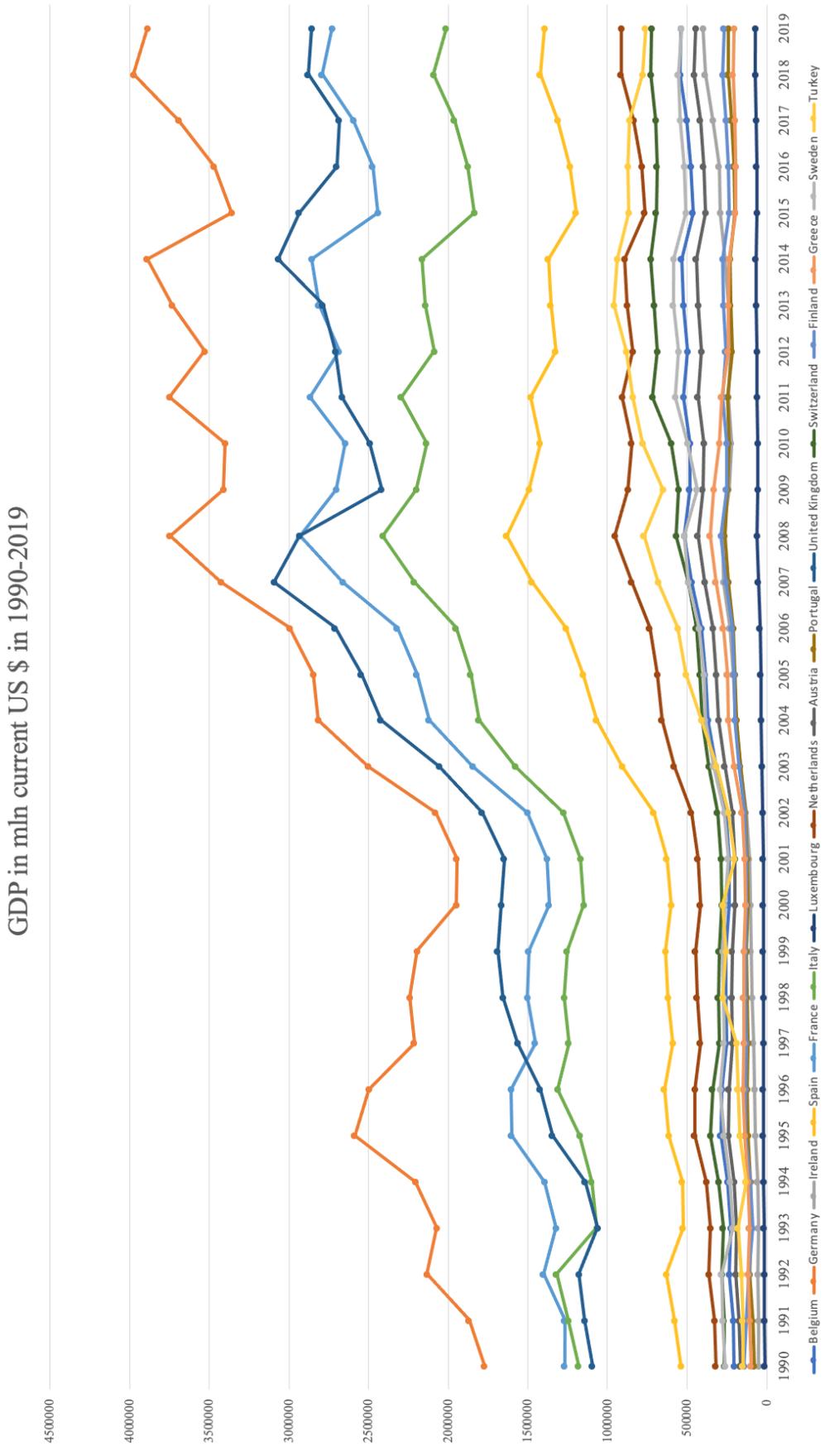

Figure 6. GDP in 1.000.000 current US $ in 1990-2019.

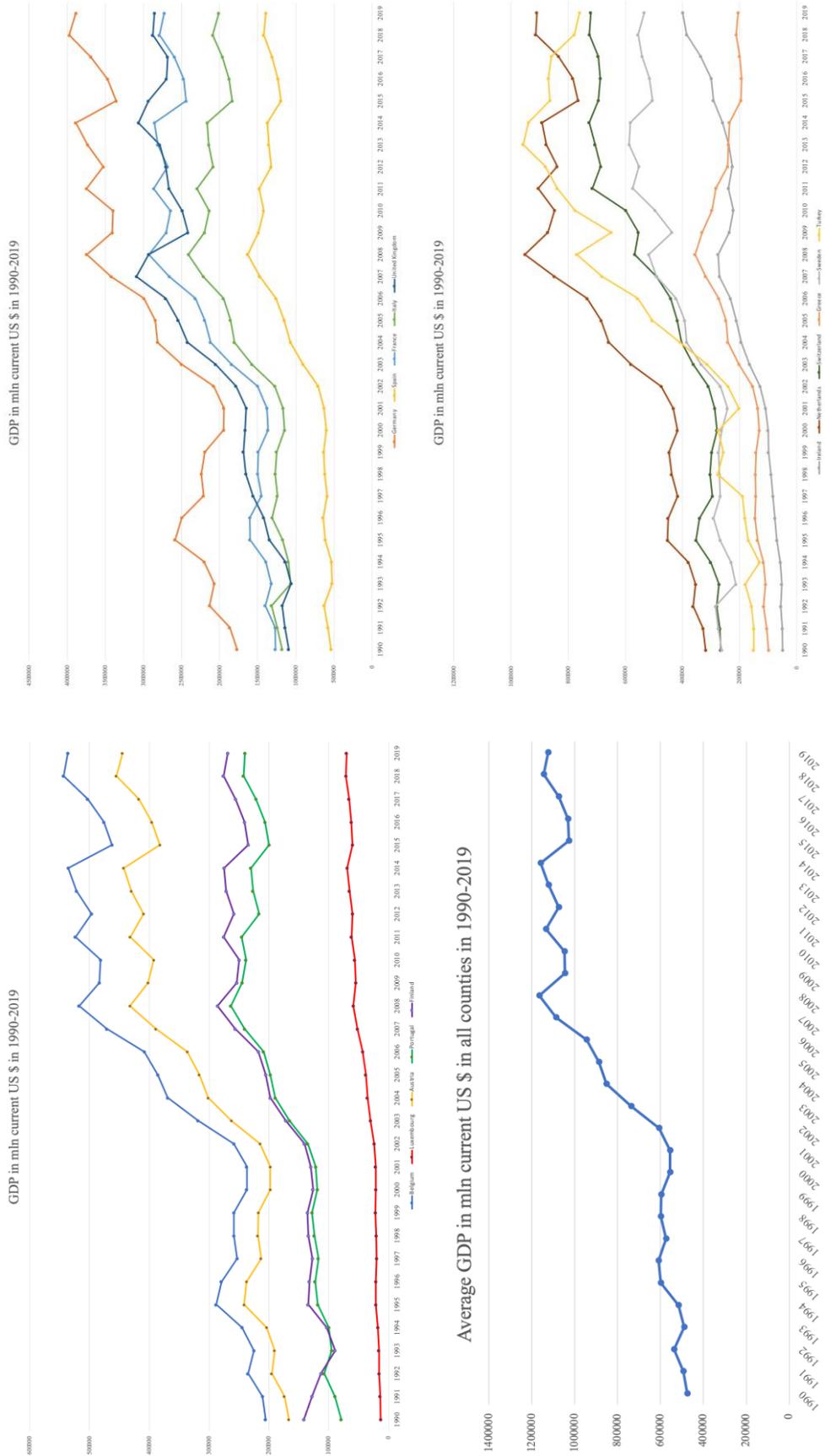

Figures 7-10. GDP in 1.000.000 current US $ in 1990-2019 and its average for all countries.

The GDP trends exhibited by the countries depicted in Figure 7 share notable similarities, with Germany standing out as the nation boasting the highest GDP over the entire period under consideration. Nevertheless, there was a discernible decline in Germany's GDP from 1996 to 2000, which can be attributed, in part, to the financial burdens incurred by the reunification of Germany and the consequential increase in expenditures directed towards supporting the Eastern regions of the country. Subsequent to this period, all the countries experienced comparable fluctuations, marked by an upswing leading up to the Global Crisis of 2007-2008, followed by a series of crests and troughs between 2008 and 2019.

Moving on to the countries enumerated in Figures 8 and 9, it is evident that they exhibit analogous patterns to the ones previously described. This phenomenon can be attributed to their geographical proximity within Europe and their majority membership in the European Union, which results in shared or interconnected economic policies. However, two outliers in this context are Turkey and Greece. Turkey, aside from the period encompassing a local economic crisis in 2001 and the Global Crisis of 2007-2008, consistently maintained an upward trajectory in its GDP, setting a noteworthy pace for the country. In 2014, this situation took a turn as the GDP declined, primarily due to a reduced economic growth rate. Greece, on the other hand, experienced an increase in GDP until 2008 but struggled to recover from the aftermath of the financial crisis, leading to a conspicuous and sustained decline in the country's GDP.

In sum, it is evident that these nations' economies are deeply interconnected, with GDP trends and passenger numbers being susceptible to the same set of external factors, including crises and political events. Notably, similar fluctuations are observable in both 2007-2008 which can be attributed to the Global Crisis.

## Weighted Mean Centres Methodology

As previously elucidated, the methodology of the Weighted Mean Gravity Centre (WMGC) proves highly conducive for scrutinizing the dynamic intricacies inherent in regional economies. This method diligently computes the central locus of a dataset, imparting due gravitas to specific data points predicated upon their pertinence and weightage in the overall context of analysis.

A salient attribute that distinguishes the Weighted Mean Centre methodology's efficacy lies in its inherent capacity to accommodate and assimilate the spatial disparities within a dataset. Through the incorporation of weighted factors, it astutely reflects the intrinsic significance attached to individual data points, thereby ensuring that the calculated centre exhibits a more nuanced and representative portrayal of the data's spatial distribution. This facet of the method assumes paramount significance when dealing with datasets characterized by variegated degrees of importance associated with their constituent elements.

The primary objective underlying the employment of the Weighted Mean Centre methodology resides in the determination of the central location or the average thereof, taking into cognizance the varying influences exerted by different data points. It achieves this objective by incorporating a multidimensional framework that accounts for spatial as well as temporal dimensions. In contrast, time-series graphical analysis, while a valuable tool in certain contexts, pales in comparison to the comprehensiveness afforded by the WMGC methodology. Time-series analysis merely delineates the temporal evolution of parameters such as passenger numbers or GDP, lacking the capability to factor in the spatial dynamics at play.

Within the purview of this paper, we adhere to the methodology endorsed by the U.S. Census Bureau for the computation of Population Centers, as expounded in their authoritative work (U.S. Census Bureau Geography Division, U.S. Department of Commerce, Washington DC 20233, 2011). This particular method takes into meticulous consideration the spheroid nature of the Earth, a planet where equatorial distances diverge from those closer to the poles. Consequently, the methodology resorts to angular degrees as the unit of measurement. To rectify the disparities in east-west distances associated with varying latitudes, the method pragmatically employs the cosine of the respective latitude to adjust the measurements. By standardizing the length of one degree along the equator as the benchmark, this approach renders east-west measurements consistent across different latitudinal positions.

For the precise calculation of longitudinal and latitudinal coordinates of population centers, the U.S. Census Bureau (U.S. Census Bureau Geography Division U.S. Department of Commerce Washington DC 20233 2011) employs the following computational formulae:

$$\bar{\phi} = \frac{\sum_i w_i \phi_i}{\sum_i w_i} \quad (1)$$

$$\bar{\lambda} = \frac{\sum_i w_i \lambda_i \cos \phi_i}{\sum_i w_i \cos \phi_i} \quad (2)$$

where $\bar{\phi}$ is the latitude of the centre of population, $\bar{\lambda}$ is the longitude of the centre of population, $\phi_i$ is a latitude of the area (region) *i*, $\lambda_i$ is a longitudee of the area (region) *i*, *w* is a population (weighting factor) in the region *i* and *i* index is a period (year).

The current paper employs a robust methodology to calculate various weighted mean centers, focusing specifically on determining the centre of economic activity and the mean gravity centres of airports. In the case of the former, we utilize the geographical coordinates of the midpoint of each country - a point equidistant from all its boundaries, readily obtainable from GPS databases (Latitude Database, 2023). To introduce a weighting factor into this calculation, we draw on the annual GDP data for each country as reported in the World Bank's World Development Indicators.

Conversely, in the context of airport mean centers, we rely on the GPS coordinates of major airports within each country. In instances where multiple airports exist, we compute the arithmetic mean of their coordinates (Airports List, 2023). The passenger data, as elaborated in preceding sections, is employed as the weighting factor in this scenario, covering the time span from 1990 to 2019.

It is imperative to clarify that the primary objective of this analysis is not the pinpoint determination of Mean Gravity Centres. This is due to the Earth's spherical nature, necessitating adjustments via the cosine function. However, these adjustments hold limited practical significance. Our foremost interest lies in comprehending the dynamic shifts of these centers, particularly with respect to cardinal directions such as North, South, North-West, and so forth (see Figure 11).

For instance, the eastward movement of airport WMGCs over time can be attributed to an increase in passengers at eastern airports or a decrease in passengers at western airports, or a combination of both factors. Therefore, the incorporation of spatial distribution data provides a richer understanding of the prevailing circumstances.

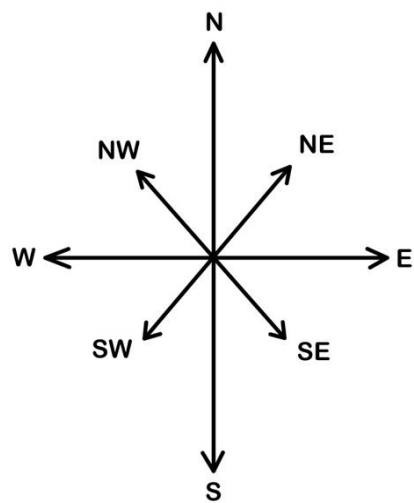

Figure 11. Cardinal directions – possible directions of MWGCs movement.

# Weighted Mean Centres Results

The results of our calculations are visually depicted through maps generated using Python and integrated with Google Maps, facilitating precise location identification of these centers. Figures 12 to 16 illustrate the shifting dynamics of airport gravity centres across the northern region of France and the Atlantic Ocean, while figures 17 to 21 capture the corresponding economic gravity centres within these same regions.

An overarching trend emerges as we analyze the data: the airport gravity centres between 1990 and 2002 closely cluster, followed by a notable shift in 2003 towards the northeast. A period of consolidation from 2012 to 2019 is evident.

Examining the early years, the trend from east to west is obvious. From 1995 to 1997, a distinct westward trajectory is observed. However, in 1998, shortly after the formation of the first airline alliance in 1997, a noticeable directional shock occurs, leading to an eastward shift. Subsequently, the directional vector fluctuates, with a westward movement between 2000 and 2002. These shifts can be attributed to the perturbations caused by the emergence of airline alliances—Star Alliance in 1997, Oneworld in 1999, and SkyTeam in 2000.

The aviation industry experiences a significant transformation in 2003, marked by a substantial directional shift, primarily eastward. This trend persists until 2011, with a predominantly consistent pace. The evolving industry landscape leads to a gradual geographical realignment of gravity centres toward central Europe. A minor deviation towards the south between 2009 and 2010 can be linked to economic fluctuations brought about by the Global Financial Crisis of 2007-2008.

The period from 2012 to 2018 witnesses considerable fluctuation in the movement of gravity centers, a phenomenon that can be attributed to economic fluctuations within the region. Overall, the centres remain closely clustered on the map.

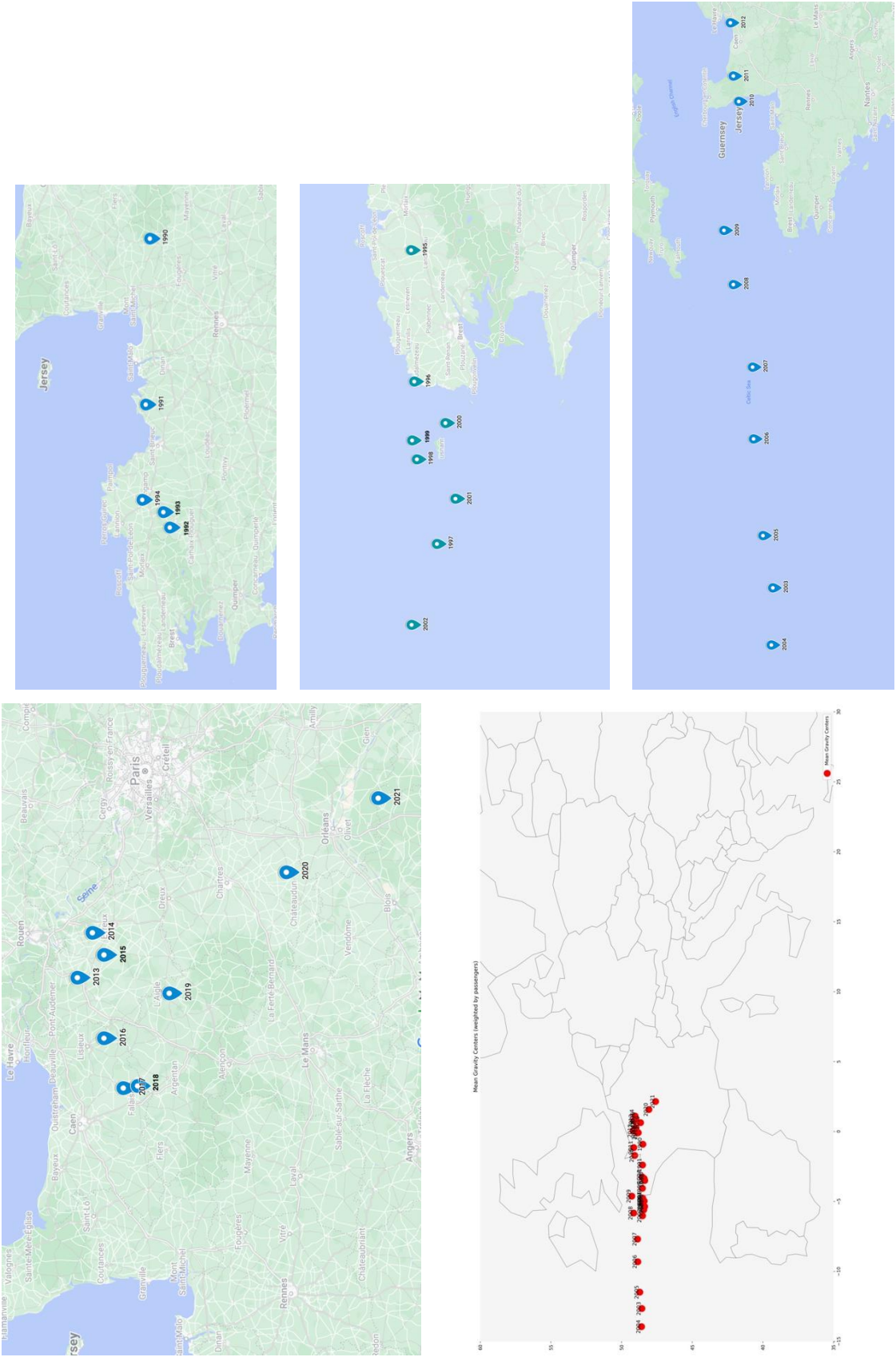

Figures 12-16. Airport weighted mean gravity centres of 1990-2019 on the map of Europe.

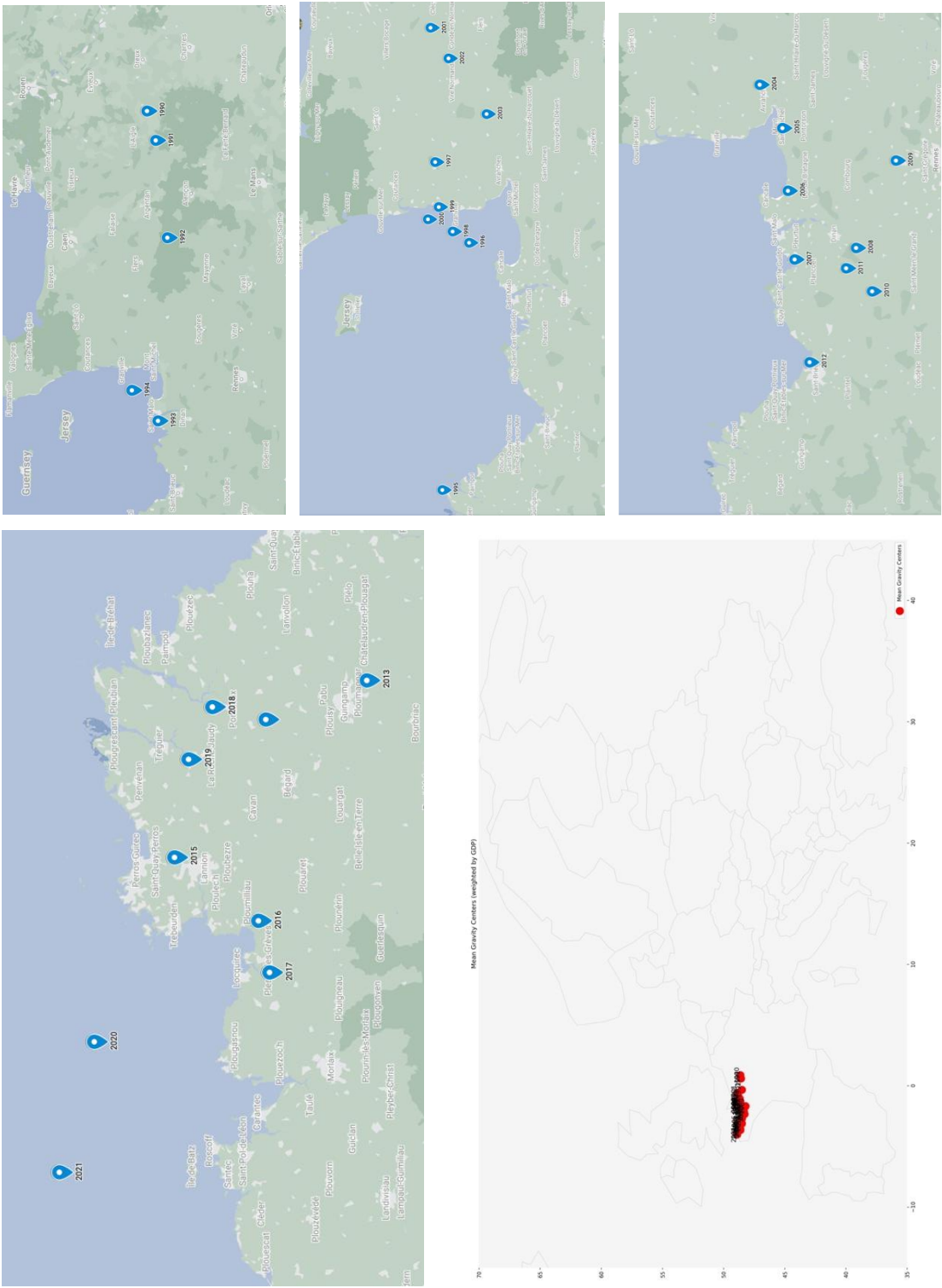

Figures 17-21. Economic weighted mean gravity centres of 1990-2019 on the map of Europe.

Economic Gravity Centers, in comparison to Airport Gravity Centers, exhibit a notable concentration on the geographical map, displaying comparatively less dynamism. The underlying rationale for these shifts can primarily be attributed to the global economic upheavals that unfolded during the specified timeframe.

In 1990-1994 the trend of movement was east to west. Between 1995 and 1996, Europe's economy underwent a phase of robust growth. Notably, the latter half of the 1990s bore witness to significant transformations in Europe. In March 1995, the signing of the Schengen Agreement marked a pivotal moment, involving the participation of seven countries. Subsequently, in 1999, the introduction of the common currency, the Euro, marked another milestone.

Starting in 2001, the Economic Gravity Centre commenced a gradual shift in a southwestern trajectory. However, the landscape changed in 2008 with the onset of the Global Financial Crisis of 2007-2008, leading to a deviation in the mean center's position. Following this deviation, there was a period of fluctuation, ultimately leading to a northwest trajectory, which prevailed until 2015.

The years from 2016 to 2019 bore witness to the transformative impact of events such as Brexit and the severe economic crisis experienced by Greece. Furthermore, Turkey, which had previously sustained remarkably high economic growth rates, transitioned towards a more moderate development trajectory. These and various other socio-economic shifts undeniably reverberated through the Economic Gravity Centres of Europe.

In sum, a correlation between Airport Gravity Centres and Economic Gravity Centres becomes apparent, as the aviation industry and regional economies are intrinsically intertwined. Nonetheless, Airport Gravity Centres exhibited more pronounced variability, influenced not only by overarching economic events such as the Global Financial Crisis of 2007-2008 but also by industry-specific occurrences, notably the formation of air alliances.

## Alliances Formation Effect

In order to assess the hypothesis regarding the impact of air alliance formation, we initially computed the standard distances from the initial geographical point (the midpoint for economic activity, and the average coordinates of airports) to the respective mean gravity centres for each year, employing the ensuing formula:

$$S = R \times arccos[\cos \phi_1 \cos \phi_2 \cos(\lambda_1 - \lambda_2) + \sin \phi_1 \sin \phi_2] \qquad (3)$$

where R is the radius of the Earth, and are the longitude and latitude for point *i* (middle point or average airport location), and are the longitude and latitude for the mean gravity centre.

In order to assess whether the establishment of air alliances had an impact on the central tendencies of airport activity and the evolving economic landscape during the observed time frame, we employed Wilcoxon test, non-parametric alternative of the t-test. The hypotheses guiding this investigation are as follows:

$H_0$: The mean of the standard deviation of the standard distances during the post-air alliance periods is equivalent to that of the pre-air alliance periods.

$H_1$: The mean of the standard deviation of the standard distances during the post-air alliance periods differs from that of the pre-air alliance periods.

Our dataset encompassed multiple years, which we categorized into three distinct groups: the first group spanned from 1990 to 1996, the second from 1997 to 2000 (the period when the airport alliances were formed), the third from 2001 to 2007, and the fourth – from 2011 to 2019. The period from 2008 to 2010 was excluded from the research because of the crisis shock, which deviates from the data and can be considered unrepresentative. Each of these groups contained an equal number of years. The results of our calculations can be found in Tables 1 and 2.

| Compared periods | P-value |
|---|---|
| *1st and 2nd* | 0.01212 |
| *1st and 3rd* | 0.01748 |
| *1st and 4th* | 0.01643 |

Table 1. T-test results for Airport weighted mean centres standard distances.

The standard deviation serves as an indicator of the rate at which passenger movements disperse. When individuals engage in frequent air travel, this indicator exhibits a heightened value. The outcomes of the analysis reveal that the establishment of air alliances has a discernible impact on the mean gravity centres pertaining to airport distribution and, in turn, the broader aviation industry. Consequently, we reject the null hypothesis ($H_0$), as there is a statistically significant difference in the means of standard deviations for standard distances between the pre-air alliance period, both post-air alliance periods and the period before the alliances formation. This implies that during the initial period, passengers displayed a lesser degree of flexibility compared to the subsequent two periods (i.e., the $2^{nd}$, $3^{rd}$ and $4^{th}$ periods).

In our study, we sought to assess the validity of our hypotheses with respect to the economic mean gravity centers. Our analysis led to the rejection of the null hypothesis ($H_0$) once again. This outcome implies that there exists a statistically significant difference in the means of standard deviations concerning standard distances between 1st and 2nd, as well as between 1st and 3rd, 1st and 4th periods. However, it is imperative to emphasize that this observed disparity should not be misconstrued as indicative of a causal relationship between the formation of air alliances and the shift in economic centers; rather, it is an empirical observation.

Furthermore, we undertook the computation of average distances from the mean airport location or the midpoint of counties to the mean gravity centre location for each year. The mathematical expression for this calculation is outlined as follows:

$$\bar{d} = \frac{1}{n}\sum_{i=1}^{n} d_i \qquad (4)$$

where *n* is the number of countries, *i* is each country's observations, *d* is a distance between mean point and the WMGC location. The results of the t-test, comparing the average of distances are presented in the Tables 3 and 4.

| Compared periods | P-value |
|---|---|
| *1st and 2nd* | 0.006061 |
| *1st and 3rd* | 0.0005828 |
| *1st and 4th* | 0.0003497 |

Table 3. T-test results for Airport weighted mean centres average distances.

For average distance measurements, analogous findings are evident. The null hypothesis, asserting that airport alliances have not influenced the placement of WMGC, is rejected. Hence, employing this methodology leads us to the substantiated conclusion that airport alliances have significantly affected the economic and transportation situation in Europe.

## Conclusions

In summary, this study endeavors to probe the correlation between the formation of air alliances and the translocation of airport gravity centres across European nations. By employing a combination of traditional descriptive analysis and the weighted mean centre methodology, the investigation furnishes valuable insights into the dynamics underpinning the distribution of airport activity and its interplay with economic factors as impacted by the inception of air alliances.

The findings unveiled in this research offer several pivotal insights. Firstly, the descriptive analysis underscores the intrinsic links between economic metrics and air passenger volumes. Remarkably, economic upheavals, such as the Global Financial Crisis of 2007-2008 and the disruptive onset of the COVID-19 pandemic, exerted profound influences on both the mean gravity centres of airports and economic activities, underscoring the intimate symbiosis between the aviation sector and the overarching economy.

Furthermore, the deployment of the weighted mean centre methodology affords a comprehensive dissection of the shifting patterns characterizing airport and economic activity. Intriguingly, the repositioning of airport gravity centres exhibited more pronounced fluctuations compared to their economic counterparts. The commencement of air alliances, commencing in 1997, emerged as a seminal factor shaping these evolutions. The outcomes of t-tests conclusively indicate that the mean standard deviations of standard distances between the periods before and after the establishment of air alliances displayed remarkable disparities for airport gravity centers. This affirms that the formation of air alliances played a substantive role in influencing the distribution of airport activities, ostensibly owing to the adjustments in route networks and operational strategies facilitated by these alliances.

Conversely, the influence of air alliances on economic gravity centers, while statistically significant in terms of mean differences, beckons further scrutiny to establish causal relationships. Economic shifts are intricately entangled within a web of multifarious determinants, encompassing global economic trends, policy shifts, geopolitical developments, and the contemporaneous emergence of air alliances.

To the best of our knowledge, this study represents a pioneering effort in utilizing geographic mean criteria as the foundation for statistical analysis of critical economic events, such as the formation of airline alliances. We illustrate that this approach may yield significant explanatory power, given the substantial structural shifts observed between the pre-alliance and post-alliance periods. These differences enable us to discern noteworthy distinctions in market conditions. Thus, we hope to inspire fellow researchers to delve further into this methodological avenue.

# List of References

# Appendix

| Year | Longitude | Latitude |
|------|-----------|----------|
| 1990 | -0,910036244 | 48,49275931 |
| 1991 | -2,393111905 | 48,51519308 |
| 1992 | -3,49611484 | 48,36623009 |
| 1993 | -3,354722794 | 48,408799 |
| 1994 | -3,245663353 | 48,53901711 |
| 1995 | -4,0607002 | 48,5223708 |
| 1996 | -4,7473138 | 48,5115999 |
| 1997 | -5,5942874 | 48,4319962 |
| 1998 | -5,1525433 | 48,50121 |
| 1999 | -5,0532226 | 48,5177836 |
| 2000 | -4,9617092 | 48,4045562 |
| 2001 | -5,3590249 | 48,3688676 |
| 2002 | -6,0184759 | 48,5210651 |
| 2003 | -12,654959 | 48,5668931 |
| 2004 | -13,945773 | 48,5902547 |
| 2005 | -11,482486 | 48,7145494 |
| 2006 | -9,3054338 | 48,8509032 |
| 2007 | -7,6865481 | 48,8667101 |
| 2008 | -5,8296882 | 49,1490869 |
| 2009 | -4,6165599 | 49,2795244 |
| 2010 | -1,7085817 | 49,0651399 |
| 2011 | -1,1457072 | 49,1422688 |
| 2012 | 0,05431244 | 49,1922279 |
| 2013 | 0,75608845 | 49,1229839 |
| 2014 | 1,10614171 | 49,0473356 |
| 2015 | 0,93303817 | 48,9897524 |
| 2016 | 0,28861698 | 48,9909014 |
| 2017 | -0,0916356 | 48,892658 |
| 2018 | -0,0756642 | 48,8199579 |
| 2019 | 0,63721638 | 48,6610601 |

Table 3. Airport mean gravity centres coordinates.

| Year | Longitude | Latitude |
|---|---|---|
| 1990 | 0,879574701 | 48,63480801 |
| 1991 | 0,596154736 | 48,58096281 |
| 1992 | -0,33118021 | 48,50976784 |
| 1994 | -2,079017664 | 48,56825742 |
| 1995 | -1,787693091 | 48,73238518 |
| 1995 | -3,0783219 | 48,8474341 |
| 1996 | -1,7494825 | 48,7522993 |
| 1997 | -1,3143245 | 48,872546 |
| 1998 | -1,6903001 | 48,8098363 |
| 1999 | -1,5583737 | 48,8587192 |
| 2000 | -1,623589 | 48,8967064 |
| 2001 | -0,5938165 | 48,888321 |
| 2002 | -0,7594141 | 48,8254464 |
| 2003 | -1,056817 | 48,6907924 |
| 2004 | -1,3006435 | 48,6592856 |
| 2005 | -1,5178975 | 48,5854539 |
| 2006 | -1,8306672 | 48,5679821 |
| 2007 | -2,175059 | 48,545604 |
| 2008 | -2,11728 | 48,3442015 |
| 2009 | -1,6796372 | 48,2117427 |
| 2010 | -2,3343639 | 48,2904095 |
| 2011 | -2,2202877 | 48,3761822 |
| 2012 | -2,6885717 | 48,497018 |
| 2013 | -3,1090662 | 48,5149767 |
| 2014 | -3,1823231 | 48,6380865 |
| 2015 | -3,4386065 | 48,7502319 |
| 2016 | -3,5570541 | 48,6471078 |
| 2017 | -3,6535344 | 48,6339967 |
| 2018 | -3,1585118 | 48,7039872 |
| 2019 | -3,2554239 | 48,7329682 |

Table 4. Economic mean gravity centres coordinates.

| Year | Averages | Standard Deviation |
|------|----------|--------------------|
| 1990 | 1486,059775 | 1337,323102 |
| 1991 | 1594,322029 | 1404,513094 |
| 1992 | 1681,045098 | 1508,595345 |
| 1993 | 1669,591612 | 1505,981523 |
| 1994 | 1660,453894 | 1507,790168 |
| 1995 | 1726,190725 | 1566,167537 |
| 1996 | 1783,254127 | 1619,087006 |
| 1997 | 1855,879844 | 1712,209982 |
| 1998 | 1817,650539 | 1681,312926 |
| 1999 | 1809,137195 | 1648,875283 |
| 2000 | 1801,679631 | 1656,698871 |
| 2001 | 1835,73201 | 1676,099324 |
| 2002 | 1892,792657 | 1720,262456 |
| 2003 | 2533,271617 | 2314,325208 |
| 2004 | 2666,653253 | 2458,842004 |
| 2005 | 2414,202189 | 2225,062154 |
| 2006 | 2198,4361 | 2033,803046 |
| 2007 | 2044,097586 | 1905,153716 |
| 2008 | 1876,117343 | 1754,854927 |
| 2009 | 1772,126611 | 1704,085541 |
| 2010 | 1541,835336 | 1590,086421 |
| 2011 | 1501,060472 | 1575,17176 |
| 2012 | 1419,15382 | 1550,00754 |
| 2013 | 1374,426966 | 1554,992981 |
| 2014 | 1352,976644 | 1573,661332 |
| 2015 | 1363,525225 | 1601,779882 |
| 2016 | 1403,986436 | 1618,600089 |
| 2017 | 1428,934398 | 1630,867594 |
| 2018 | 1428,040355 | 1629,044859 |
| 2019 | 1382,612506 | 1592,292041 |

Table 5. The standard deviation and averages of standard distances for Airport mean gravity centres coordinates in km.

| Year | Averages | Standard Deviation |
|------|----------|-------------------|
| 1990 | 1484,042642 | 1334,87712 |
| 1991 | 1499,441404 | 1342,934256 |
| 1992 | 1552,657689 | 1380,617059 |
| 1993 | 1662,001131 | 1499,089273 |
| 1994 | 1642,899458 | 1441,612887 |
| 1995 | 1729,19054 | 1548,759044 |
| 1996 | 1640,432368 | 1463,284501 |
| 1997 | 1612,781374 | 1440,888906 |
| 1998 | 1636,639538 | 1506,42839 |
| 1999 | 1628,22844 | 1484,471522 |
| 2000 | 1632,436305 | 1506,343903 |
| 2001 | 1568,41978 | 1396,666655 |
| 2002 | 1578,367931 | 1419,324392 |
| 2003 | 1596,596772 | 1454,97337 |
| 2004 | 1611,848243 | 1487,574278 |
| 2005 | 1625,669067 | 1527,245845 |
| 2006 | 1645,795676 | 1554,463685 |
| 2007 | 1668,358044 | 1591,188439 |
| 2008 | 1665,003424 | 1606,542371 |
| 2009 | 1636,978363 | 1564,981758 |
| 2010 | 1679,546818 | 1647,884408 |
| 2011 | 1671,711368 | 1639,779157 |
| 2012 | 1702,79679 | 1690,605311 |
| 2013 | 1731,570374 | 1731,314622 |
| 2014 | 1736,470874 | 1715,017596 |
| 2015 | 1754,305629 | 1731,699245 |
| 2016 | 1762,738621 | 1748,905438 |
| 2017 | 1769,598257 | 1745,368927 |
| 2018 | 1734,762045 | 1666,776736 |
| 2019 | 1741,48575 | 1668,203811 |

Table 5. The standard deviation of standard distances for Economic mean gravity centres coordinates in km.